\def\bk{{\bf k}}

\def\tu{\tilde{u}}
\def\tv{\tilde{v}}
\def\tw{\tilde{w}}
\def\tit{\tilde{t}}

\def\bea{\begin{eqnarray}}
\def\eea{\end{eqnarray}}
\def\ba{\begin{array}{l l}}
\def\ea{\end{array}}

\documentclass[epj]{svjour}
\usepackage{epsfig}
\usepackage{amssymb,amsmath}

\begin{document}
\numberwithin{equation}{section}

\title{Quantum tricriticality  in transverse Ising-like systems}

\author{M.~T. Mercaldo\inst{1,2} \and  I. Rabuffo\inst{1,2} \and A. Naddeo\inst{1,2} \and A. Caramico~D'Auria\inst{3} \and L. De~Cesare\inst{1,2}}
\institute{ Dipartimento di Fisica ``E. R. Caianiello'',
Universit\`a di Salerno ,  I-84084
Fisciano (Salerno), Italy \and
CNISM, Unit\`a di Salerno, I-84084
Fisciano (Salerno), Italy \and
Dipartimento di Scienze Fisiche,
Universit\`a di Napoli Federico II, I-80125
Napoli, Italy  }

\abstract{
The quantum tricriticality of $d$-dimensional transverse
Ising-like systems is studied by means of a perturbative
renormalization group approach focusing on static susceptibility.
This allows us to obtain the phase diagram for
$3\leq d<4$, with a clear location of the critical lines ending in
the conventional quantum critical points and in the quantum
tricritical one, and of the tricritical line for temperature
$T\geq 0$. We determine also the critical and the tricritical
shift exponents close to the corresponding ground state
instabilities. Remarkably, we find a tricritical shift exponent
identical to that found in the conventional quantum criticality
and, by approaching the quantum tricritical point increasing the
non-thermal control parameter $r$, a crossover of the quantum
critical shift exponents from the conventional value $\phi  =
1/(d-1)$ to the new one $\phi  = 1/2(d-1)$. Besides, the
projection in the $(r,T)$-plane of the  phase boundary
ending in the quantum tricritical point and crossovers in the
quantum tricritical region appear quite similar to those found
close to an usual quantum critical point. Another feature of
experimental interest is that the amplitude of the Wilsonian
classical critical region around this peculiar critical line is
sensibly smaller than that expected in the quantum critical
scenario. This suggests that the quantum tricriticality is
essentially governed by mean-field critical exponents,
renormalized by the shift exponent $\phi  = 1/2(d-1)$  in the
quantum tricritical region.
}
\maketitle

\section{Introduction}
\label{sec1}

The last decades have seen a renovated interest in the topic of
quantum phase transitions (QPTs), which have had a surprising
impact in explaining several exotic low-temperature properties of
a variety of innovative materials \cite{sachdev,Cont,Stew,Lohn,Gege}
which do not  fit conventional many-body theories.
Moreover, even though a QPT is strictly
speaking a zero-temperature instability, its experimental
manifestations are clearly observed at finite temperature, within a
rather wide region around the quantum critical point
(QCP) \cite{Lohn,Gege}.

Recent experimental \cite{ZrZn,UGe1,UGe2,MnSi} and theoretical \cite{schm04,ted05} analysis of ferromagnetic  QPTs
have shown  that the second-order phase transition turns into a
first-order one when approaching the absolute zero temperature.
Thus, the emerging scenario indicates the presence of a
tricritical point in proximity of a QPT, that
prevents the long-range order to be continuously connected with
the disordered phase when the quantum fluctuations become
dominant.

In principle, it is possible to tune opportunely the control parameters in order
to move the tricritical point to zero
temperature, creating then a quantum tricritical point (QTCP). The
previous considerations
pose the problem on a broader context and point
the interest to the nature of the critical fluctuations around a
QTCP, as well as about their possible physical implications.

Indeed, there are experimental evidences for some compounds
suggesting that they are close to a QTCP. Examples are
given by the oxides Sr$_3$Ru$_2$O$_7$ \cite{grig01,grig03,green05}
and Pr$_2$CuO$_4$ \cite{pla03}, the heavy fermions URhGe
\cite{hux07} and  YbRh$_2$Si$_2$ \cite{misa08,misa09}, and the
pnictide superconductor LaFeAsO \cite{ironsc}.
Let us present some more  detail for some of the mentioned systems. Concerning
 the ruthenate compound Sr$_3$Ru$_2$O$_7$, a metamagnetic QCP was found \cite{grig01}, where the field angle $\theta$
(measured with respect to the ab plane) acts as a tuning
parameter, allowing the construction of a $(H,\theta,T)$ phase
diagram, where $H$ is the magnetic field and $T$ is the
temperature, for metamagnetism and quantum criticality
\cite{grig03}. The metamagnetic transition tracks a first-order
line in the $(H,\theta)$-plane which terminates at a QCP when the temperature, and thus the energy, scale
associated with the metamagnetic transition is tuned to zero.
Successive studies \cite{green05} on ultraclean samples of
Sr$_3$Ru$_2$O$_7$ have revealed striking anomalies in magnetism
and transport in the vicinity of the QCP. In particular,
measurements of resistivity and ac magnetic susceptibility have
shown an intricate phase diagram where the line of first-order
metamagnetic transition in the $(H,\theta)$ plane appears to
bifurcate in two transition lines (observed for $H||c$). Such a complex
phase behavior can be ascribed to a ``symmetry-broken" tricritical
point structure \cite{green05}. Thus the bifurcation mechanism
seems to provide an opportunity to realize quantum tricritical
behavior in Sr$_3$Ru$_2$O$_7$.
This feature seems to happen also in the heavy-fermion compound URhGe \cite{aoki01}.
This material has been intensively studied due to
the emergence of superconductivity close to its ferromagnetic QCP.
Detailed experimental analysis of the magnetic field vs. temperature phase
diagram show,  at very low temperatures, the presence of a
tricritical point at which the continuous transition line
bifurcates \cite{hux07}. The superconducting phase in URhGe occurs
around this bifurcation close to zero temperature. This situation
suggests that URhGe is close to a QTCP and
not just an ordinary QCP \cite{hux07}, but it has not
so far been examined theoretically. Studies in this sense could give new
interesting information about the mechanism driving the
unconventional superconductivity.
 Finally, very recently it has been speculated
that also  iron pnictide superconductors are close to a
QTCP \cite{ironsc}.
Neutron scattering experiments \cite{cruz} have shown that lowering temperature the layered parent compound LaFeAsO
first undergoes a structural phase transition, with a distortion from tetragonal structure to monoclinic structure,
followed by a spin-ordering phase transition.
Upon doping, magnetic order
is suppressed and disappears at zero temperature (QCP).
The finite-temperature magnetic transition shows basically two types of behavior \cite{krell,jesche}:
in some compounds (particularly of the 122 family, such as SrFe$_2$As$_2$)
the transition is first order, while in others (the 1111 family, such as LaFeAsO) it appears second order, thus suggesting that
these materials  are close to a tricritical point. Hence, it has been speculated\cite{ironsc} that opportunely
tuning two non-thermal control parameter (like pressure and doping) the tricritical point can be driven to zero
temperature producing a QTCP.

Theoretically the subject of
classical tricritical behavior  (i.e. when the tricritical point
is located at finite temperature) is well known \cite{classTC}. Much less
has been done, instead, about quantum tricriticality, i.e. exploring  the
low-temperature properties and crossovers around a QTCP.
The first study  of quantum tricritical phase transitions
was done via renormalization group (RG) theory \cite{bus}, for a variety of quantum systems with and without
quenched disorder, but it was  carried out  only at $T=0$, although the general temperature-dependent flow equations were presented. Implications
at finite temperature in the influence domain of a QTCP were actually missing. In the last years other few
theoretical works appeared on subject related to quantum
tricriticality. Schmalian and Turlakov \cite{schm04} have considered the
QPTs of magnetic rotons, in order to explain
part of the phenomenology of the itinerant ferromagnet MnSi. Using the
self-consistent Hartree approximation, they studied weak
fluctuation-driven first order QPTs. For
certain values of the parameters of the model under study they
found a QTCP with accompanied non-Fermi
liquid behavior of the electrons. In Refs. \cite{misa08,misa09}
 quantum tricriticality is analyzed by means of
self-consistent renormalization theory \cite{moriya} for itinerant
antiferromagnets. The authors suggest that the presence of a
QTCP can explain the otherwise puzzling coexistence of ferromagnetic and antiferromagnetic fluctuations
observed in the heavy fermion compound YbRh$_2$Si$_2$.
More recently, a functional RG analysis of quantum triciticality in metals has been
presented \cite{metz}, where the correlation length has been evaluated
in the quantum tricritical region, also at low-but finite temperature.
Thus we clearly see that recently the relevance of quantum tricriticality has been emphasized in
several  theoretical and experimental works \cite{schm04,ted05,green05,hux07,misa08,misa09,ironsc,metz},
as quantum tricritical fluctuations could give a new route to explain some of
the puzzling phenomena occurring at QPTs.

In this paper, using the conventional RG approach  to quantum critical
phenomena \cite{hertz,millis}, we analyze the influence at finite
temperature of the QTCP and we study several crossovers occurring in the
phase diagram, mainly focusing on the behavior of static
susceptibility. To tackle this problem
 we consider  the $n$-vector  ($O(n)$) generalization of the quantum action \cite{sachdev} for the
transverse Ising model (TIM), which is one of the prototypical examples capturing the essence of QPTs,
to include a variety of quantum systems which may exhibit a QTCP. We refer to these as ``transverse Ising-like systems" (TIM-like systems).
The TIM was  originally
employed to model the order-disorder transition in some double-well ferroelectric systems such as
potassium dihydrogen phosphate (KH$_2$PO$_4$) crystals \cite{blinc60,kat62,degen63},
 and since it has been successfully applied to
several other materials as ferromagnets with a strong uniaxial
anisotropy and further ferroelectric compounds with quantum
structural phase transitions.
One of the best example of experimental realization of TIM is given by the family of compounds
 LiREF$_4$ (where RE stands for rare earth). In particular experimental analysis of susceptibility for
the insulating magnet LiHoF$_4$ shows that the quantum critical behavior
is mean-field like \cite{aeppli}, as expected by conventional theory on TIM.
Other magnetic materials, whose QPT may be described using TIM, are some compounds belonging to the RE(OH$_3$) family \cite{stas08}.
An extensive discussion of the
applications of TIM and its extensions can be found in
Refs. \cite{sachdev,stinch,chakra1,chakra2,dutta}.
Moreover, recent theoretical investigation on magnetic orders and lattice structure in
iron-based superconductors \cite{XU1,XU2} have shown that for high doping and low pressure
the spin-density-wave transition
belongs to the $O(3)$ tridimensional  universality class,
with dynamical critical exponent $z=1$, which enters the subject of our analysis.
Hence our work  may be also of some interest for these systems,
where indeed the existence of a QTCP  has been speculated \cite{ironsc}.

The paper is organized as follows. In Sec.\ref{sec2} we present
the model and the RG framework, and we derive the low temperature
phase diagram. In Sec.\ref{sec3} we describe the quantum
tricriticality and crossovers in the influence domain of a QTCP,
through the analysis of the static susceptibility. Finally in
Sec.\ref{sec4} some conclusions are drawn.

\section{The model and the renormalization group approach}
\label{sec2}

A remarkable feature in theory of critical phenomena is universality,
 so that through a reduced number of models
one can describe a wide variety of systems. In particular for QPTs
a given universality class is defined by the space dimensionality
$d$, the order parameter symmetry index $n$ and the dynamical
critical exponent $z$ that characterizes the intrinsic dynamics of
the system. In this work we consider a $n$-vector Ginzburg-Landau
 action, which in the Fourier space is written in the form \cite{misa08,misa09,bus,metz}
 (in convenient units)
\bea +\label{S1} && S\{\psi\} = \frac12
\sum_{j=1}^n\sum_{\bk,\omega_l}
(r_0+k^2+f(\bk,\omega_l))|\psi^j(q)|^2  \\ \nonumber &&+ \frac{u_0
T_0}{4 V} \sum_{i,j=1}^n \sum_{q_1,q_2,q_3} \psi^i(q_1)
\psi^i(q_2) \psi^j(q_3)\psi^j(-q_1-q_2-q_3)\\ \nonumber &&+
\frac{v_0 T_0^2}{6 V^2} \sum_{i,j,l=1}^n \sum_{\{q_\nu\}}
\delta_{\sum_{\nu=1}^6 q_\nu,0} \psi^i(q_1) \psi^i(q_2)
\psi^j(q_3)\\ \nonumber &&\qquad \times
\psi^j(q_4)\psi^l(q_5)\psi^l(q_6), \eea
where $\psi^j(q)$ are the
Fourier components of an $n$-vector real order parameter field, $q
\equiv (\bk,\omega_l)$, $\bk$ is the wave vector with cutoff
$\Lambda$, $\omega_l=2 \pi l T_0$ ($l=0,\pm1,\pm2,..$) are bosonic
Matsubara frequencies, $T_0$ is the temperature and $V$ is the
volume. In order to study tricritical behavior, we need to keep an
expansion of the order parameter field up to sixth order. In the
action \eqref{S1} the function $f(\bk,\omega_l)$, which defines
the intrinsic dynamic, and the parameters $r_0, u_0$ and $v_0$
depend on the physical system under study.
In this paper, as mentioned in the introduction,  we focus on TIM-like systems for which
 $f(\bk,\omega_l) = \omega_l^2$ and the dynamical critical exponent is $z=1$.
However, our analysis is quite general and may be easily extended to
other quantum systems with intrinsic dynamic described by
$f(\bk,\omega_l)=|\omega_l|^\mu/k^{\mu'}$ $(\mu\geq1,\mu'\geq0)$ in the
quantum action and with a dynamical critical exponent $z=(2+\mu')/\mu$ \cite{nota2}.

At mean field level (i.e. in absence of fluctuations)
the considered  model yields a
continuous phase transition for $u_0>0$ and $r_0=0$, a first-order transition for $u_0<0$ and $r_0=3 u_0^2/16 v_0$, and a
tricritical point for $u_0=0$ and $r_0=0$, while the parameter $v_0$ has to be  positive for thermodynamic stability.

To analyze the critical properties of the model $\eqref{S1}$, taking into account the effects of fluctuations,
we employ the momentum-shell RG  procedure which gives a flow of
Hamiltonians whose interaction parameters satisfy a set of coupled nonlinear
differential equations \cite{bus}
\begin{subequations}
\label{rg1}
\bea
\frac{dr}{dl} &=& 2 r + (n+2) K_d F_1(r,T) u\;,\\
\frac{du}{dl} &=& (4-(d+z))u -(n+8)K_d F_2(r,T) u^2+ \nonumber  \\&+& (n+4)K_d F_1(r,T) v \;,\\
\frac{dv}{dl} &=& 2(3-(d+z))v-3(n+14)K_d F_2(r,T) u v\;,\\
\frac{dT}{dl} &=& z T \label{teq}
\eea
\end{subequations}
where $l$ is the scaling parameter, $K_d=[2^{(d-1)} \pi^{d/2}\Gamma(d/2)]^{-1}$ and
$F_n(r,T) = T\sum_{\omega_l}[r+1+f(1,\omega_l)]^{-n}$,  i.e. explicitly for the systems we are analyzing with $f(\bk,\omega_l)=\omega_l^2$ and $z=1$
\bea
F_1(r,T) &=& \frac{\coth(\sqrt{1+r}/2T)}{2\sqrt{1+r}} \;,\\
F_2(r,T) &=& \frac{\coth(\sqrt{1+r}/2T)}{4(1+r)^{3/2}}+\frac{\rm{csch}^2(\sqrt{1+r}/2T)}{8 T\sqrt{1+r}}\;.
\eea
Equation \eqref{teq} can be solved immediately yielding
\bea
T(l) = T_0 {\rm e}^{ l }\;.
\eea

Following the standard procedure \cite{millis}, we need to separate the low and high temperature regimes, so that for the
quantum regime, with $T(l)\ll 1$  (which includes the case $T=0$), Eqs.\eqref{rg1} become
\begin{subequations}
\label{rgT0}
\bea
\frac{dr}{dl} &=& 2 r + \frac{n+2}2 K_d \frac{u}{\sqrt{1+r}}\;,\\
\frac{du}{dl} &=& (3-d)u -\frac{n+8}4 K_d \frac{u^2}{(1+r)^{3/2}}+ \nonumber \\&+& (n+4)K_d \frac{v} {\sqrt{1+r}}\;,\\
\frac{dv}{dl} &=& 2(2-d)v-\frac34(n+14)K_d\frac{ u v}{(1+r)^{3/2}}\;,
\eea
\end{subequations}
except for exponentially small corrections in $T(l)$. Equations \eqref{rgT0} are not
suitable to study quantum tricritical behavior for
$d<2$ since, as in the corresponding classical framework \cite{stephen}, higher order
contributions to flow equations become necessary.

At finite temperature or classical regime, with $T(l)\gg 1$, the recursion equations are
\begin{subequations}
\label{rgT}
\bea
\frac{dr}{dl} &=& 2 r + (n+2) K_d \frac{\tu}{1+r}\;,\\
\frac{d\tu}{dl} &=& (4-d)\tu -(n+8) K_d \frac{\tu^2}{(1+r)^{2}} + \nonumber \\&+& 2 (n+4)K_d \frac{\tv} {1+r}\;,\\
\frac{d\tv}{dl} &=& 2(3-d)\tv-3(n+14)K_d\frac{ \tu \tv}{(1+r)^{2}}\;,
\eea
\end{subequations}
where the appropriate coupling parameters are $\tu = T u$ and $\tv = T^2 v$.

These two regimes are separated by
the condition $T(\bar{l})\simeq 1$ at a given value $\bar{l}$ of the rescaling parameter $l$.
 To explore the finite temperature critical behavior we must solve the set of differential
equations \eqref{rgT}, using as initial conditions the solution of the set of equations
\eqref{rgT0} evaluated in $\bar{l}$.
In the remaining part of the paper we will work for dimensionalities $d\geq 3$, where Eqs.\eqref{rgT} are suitable.

\subsection{Solution of the flow equations}

Let us first consider the quantum regime ($T(l)\ll 1$).
To solve the flow equations \eqref{rgT0} it is convenient
to introduce two new coupling parameters, which are  linear combinations of the original ones,
\bea
t&=&r+\frac{n+2}{2(d-1)}K_d\left(u+\frac{n+4}4 K_d \frac v{d-1}\right)\;,\\
w&=&u+\frac{n+4}{2(d-1)}K_d v\;,
\eea
 so that the equation for $w$ (to leading order in the coupling parameters) is decoupled from the other equations.
Then, the general solution of the flow equations in the low-$T$ regime, in terms of these new parameters, is given by
\begin{subequations}
\label{qreg}
\bea
t(l) &=& t(0) e^{2l} Q(l)^{-(n+2)/(n+8)} \label{tl}\;,\\
w(l) &=& w(0) e^{(3-d) l} Q(l)^{-1}\;,\\
v(l) &=& v(0) e^{2(2-d)l} Q(l)^{-3(n+4)/(n+8)}\;,
\eea
\end{subequations}
where $Q(l) = 1+K_d (n+8)w(0)(e^{(3-d) l}-1)/(4(3-d))$ and
\begin{subequations}
\label{incond}
\bea
t(0)&=&r_0+\tfrac{(n+2)K_d}{2(d-1)}\left(u_0+\tfrac{(n+4)}{4(d-1)}K_d v_0  \right)\equiv t_0\,,\quad\\
w(0)&=&u_0+\frac{n+4}{2(d-1)}K_d v_0 \equiv w_0\;,\\
v(0)&=&v_0\;.
\eea
\end{subequations}
The  expressions \eqref{qreg} may be simplified since, working for
dimensionalities $d \gtrsim 3$, we have $Q(l)\simeq 1$ for $l\gg 1$.
This is equivalent to linearizing the RG equations around the gaussian fixed point (GFP) , which is the one that governs tricriticality.
Basically, we have that fluctuations shift the conditions for tricriticality given by Landau theory from $(r_0=0,u_0=0)$ to $(t_0=0,w_0=0)$.
Hence we have  continuous phase transitions for $t_0=0$ and $w_0>0$ (or $u_0 > - (n+4)K_d v_0/2(d-1)$ so that also
negative values of $u_0$ are allowed) and a tricritical point for $t_0=0$ and $w_0=0$.
For $w_0<0$ there are no stable fixed points, so nothing can be said in this region within the RG approach,
but matching with Landau theory we can assume that first order phase transitions occur.
Restoring the original parameters at $T_0=0$ (from the conditions $t_0=w_0=0$) we obtain the coordinates of the QTCP
\bea
\label{QTCP}
r_{tr}= \frac{(n+2)(n+4) K_d^2 v_0}{8(d-1)^2} \;,\quad u_{tr}=-\frac{(n+4)K_d v_0}{2(d-1)}\;.\qquad
\eea
We stress that fluctuations shift the coordinate $u_0=0$ of the tricritical point, as predicted from Landau theory,
to the lower value $u_0= u_{tr}(<0)$.
We also determine the line of QCPs in the $(r_0,u_0)$-plane
\bea
\label{QCPline}
r_{c} = -\frac{n+2}{2(d-1)}K_d \left(u_0 + \frac{n+4}{4(d-1)}K_d v_0\right)\;,
\eea
parameterized by $v_0>0$. This equation defines a straight line which ends at the QTCP
with $r_{c}$ increasing from negative to positive values up to the  QTCP coordinate $r_{c}=r_{tr}>0$,
by variation of the parameter $u_0$ from positive to negative values provided that $w_0\geq 0$.

We now provide an estimate of the line  separating the classical and quantum regime, i.e. when $T(\bar{l})\simeq 1$,
focusing on the scaling field $t(l)$ which contains the relevant information about the structure of the phase
diagram and the physics of the systems. We know that one can stop the scaling when $t(l)\sim 1$ \cite{millis,rudne},
hence with the approximate solution $t(l)\simeq t_0e^{2l}$  close to the GFP, we obtain for the rescaling parameter
a value $l^*\gg 1$ ($e^{l^*}=t_0^{-1/2}$) which can be used  to evaluate the renormalized temperature $T(l)$, requiring that $T(l^*)\ll 1$ in order
to stay within the quantum regime. This gives us the condition $T_0 \ll t_0^{1/2}=(r_0-r_c)^{1/2}$ for the occurrence of this asymptotic regime.
In particular, 
if $u_0=u_{tr}$ we have $T_0 \ll (r_0-r_{tr})^{1/2}$. Hence
there is a crossover line, separating quantum  ($T(l)\ll 1$) and classical ($T(l)\gg 1$) regimes for $r_0>r_{tr}$, which is given by
\bea
\label{QR}
T^\dagger \simeq (r_0-r_{tr})^{1/2}\;.
\eea

As mentioned before, to study the low-temperature phase diagram we have
to solve the set of coupled differential equations \eqref{rgT} appropriate at finite temperature,
where we assume as initial conditions the solution \eqref{qreg} of the quantum regime
evaluated at $\bar l = \ln (1/T_0)$.

Also in the classical regime it is convenient to rewrite the recursion relations by introducing two
new coupling parameters
\bea
\tit&=&r+\frac{n+2}{d-2}K_d\left(\tu+\frac{n+4}{d-2} K_d \tv\right)\;,\\
\tw&=&\tu+\frac{2(n+4)}{d-2}K_d \tv\;,
\eea
in terms of which the solution of Eqs.\eqref{rgT} is
\begin{subequations}
\label{creg}
\bea
\tit(l) &=& \tit(\bar{l}) e^{2(l-\bar{l})} \tilde{Q}(l,\bar{l})^{-(n+2)/(n+8)}, \label{tdielle}\\
\tw(l) &=& \tw(\bar{l}) e^{(4-d) (l-\bar{l})} \tilde{Q}(l,\bar{l})^{-1},\\
\tv(l) &=& \tv(\bar{l}) e^{2(3-d)(l-\bar{l})} \tilde{Q}(l,\bar{l})^{-3(n+4)/(n+8)}\;,
\eea
\end{subequations}
where $\tilde{Q}(l,\bar{l}) = 1+(n+8)K_d \tw(\bar{l})(e^{(4-d) (l-\bar{l})}-1)/(4-d)$.
The results in the classical regime are very similar to those
obtained in Ref. \cite{aha}. For $3\lesssim d<4$ there is a GFP which  is doubly
unstable having two relevant scaling fields, $\tit(l)$ and $\tw(l)$, and a critical  FP which is unstable only against the
scaling field  $\tit(l)$. Thus, when $\tit(\bar{l})=0$ with $\tw(\bar{l})>0$,
the Hamiltonian flow evolves towards the critical FP, yielding a continuous
phase transition, while for $\tw(\bar{l})<0$ the critical  FP is no longer accessible.
Therefore, the conditions $\tit(\bar{l})=0$ and $\tw(\bar{l})=0$ define a finite-temperature tricritical point, marking
the borderline between these two regimes. The dependence on $\bar{l}$ implies a dependence on
temperature $T_0$ (from now on we will omit the subscript 0 for the physical temperature  and for the other coupling parameters, too). Hence
the condition $\tit(\bar{l})=0$ with $\tw(\bar{l})>0$ determines the equation for the critical surface at finite temperature,
while $\tit(\bar{l})=0$ with $\tw(\bar{l})=0$ provides the finite-temperature tricritical line.

\subsection{Critical and tricritical lines and low temperature phase diagram}

Using the solution of the RG equations derived in previous subsection and the conditions for the relevant scaling fields
we obtain the second-order phase transition surface in the parameters space
\bea
\label{supcrit}
r_c(T) = r_c-a_{n,d} \left(u -u_{tr}\right) T^{d-1} - b_{n,d} v T^{2(d-1)}\,,\quad
\eea
under the condition for $u$
\bea
\label{utr}
u&>& -\frac{n+4}{2(d-1)} K_d v \left(1+\frac{3d-2}{d-2} T^{d-1}\right) \\ \nonumber
&=&  u_{tr}\left(1+\frac{3d-2}{d-2} T^{d-1}\right)\;,
\eea
where $r_c$ (which explicitly depends on $u$ and $v$) is given by Eq.\eqref{QCPline} and
\bea
a_{n,d} &=& \frac{d(n+2) K_d}{2(d-1)(d-2)}\;,\\
b_{n,d} &=& \frac{5d^2-8d+4}{8(d-1)^2(d-2)^2}(n+2)(n+4) K_d^2\;.
\eea
For fixed values of $u > u_{tr}$ and for $v>0$, we obtain a family of critical lines in the $(r,T)$-plane, ending in QCPs.
At sufficiently low temperature close to the QCP
\bea
r_c(T) = r_c - a_{n,d}(u-u_{tr}) T^\psi,
\eea
where  $\psi = d-1$. In this way we recover the usual shift exponent for systems in the universality class
which is the subject of this paper. Increasing the temperature, a different $T$-behavior with exponent $2\psi$ governs
$r_c(T)$. Then, a crossover temperature $T^*$ exists separating
two distinct asymptotic $T$-behaviors of the phase boundary, with power-laws $\psi$ for $T\ll T^*$ and $2\psi$ for $T\gg T^*$, with
\bea
T^* =\left[\frac{a_{n,d}(u-u_{tr})}{b_{n,d} v}\right]^{1/\psi}\;.
\eea

When $u = u_{tr}$, among the  family of critical lines in the $(r,T)$-plane, parameterized by $u$ and $v$,
we select the one ending in the QTCP, characterized by the temperature behavior
\bea
r_{{qtc}}(T) = r_{tr} - b_{n,d} v T^{2 \psi}\;,
\eea
where the index {\it qtc} stands for ``quantum tricritical".

In terms of temperature, this peculiar phase boundary is determined by the equation
\bea
T_{qtc}(r)=\left(\frac{r_{tr}-r}{b_{n,d}v}\right)^{1/2\psi}\qquad ({\rm for}\;\; r<r_{tr})\;,
\eea
which provides the shift exponent $\phi=1/2\psi=1/2(d-1)$, to be compared with $\phi=1/\psi=1/(d-1)$ which occurs when
the critical lines ends in an ordinary QCP. Of course as  $d\to 3^+$, we should have $\phi=1/2$ close to
a QCP for $T\ll T^*$ and $\phi=1/4$ close to the QTCP.

\begin{figure}
  \begin{center}
   \includegraphics[width=8cm]{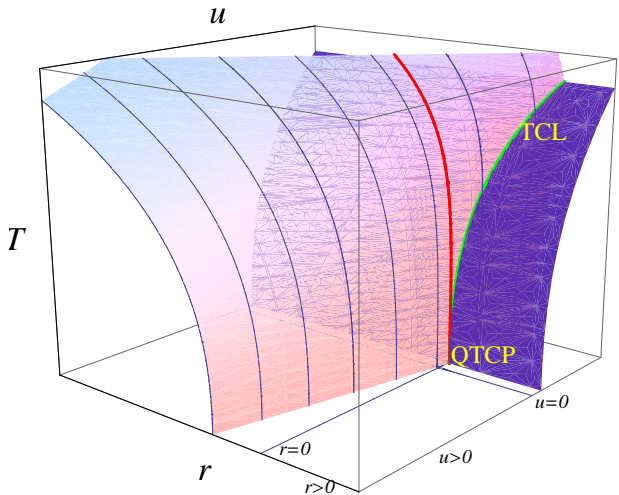}
  \end{center}
  \caption{Schematic phase diagram in the $(r,u,T)$-space for $n\geq 1$ and $3\lesssim d<4$. The critical surface is the light one and
  the lines on it are critical lines ending in ordinary QCPs for different fixed values of
  $u>u_{tr}$, while the thicker line (red online) is the one ending in the
  QTCP, named $r_{qtc}(T)$ in the text. The dark surface in the back is given by $\tw(\bar{l})=0$ (see text).
  In the region of the phase diagram behind this
  surface first-order phase transitions are expected to occur.  The intersection of the two surfaces (green online)
  is the finite-temperature tricritical line $r_{tcl}(T)$.}
 \label{fig1}
\end{figure}

From the conditions $\tit(\bar{l})=0$ and $\tw(\bar{l})=0$, as written above,
we get  the expression of the finite-temperature tricritical line (TCL)
\bea
\label{TCL}
r_{{tcl}}(T) = r_{tr}\left[1+\frac{2(3d-2)}{d-2} T^{\psi}+O(T^{2\psi})\right]\;.
\eea
If we consider the projection of the TCL in the $(r,T)$-plane we obtain a tricritical shift exponent given by $\phi_{tr}=1/\psi=1/(d-1)$,
which is the same value we obtain for conventional criticality. In particular
for $d\simeq 3$ we have $\phi_{tr}=1/2$,

In Fig. \ref{fig1}  we present the schematic low-$T$ phase diagram in the ($r,u,T$)-space, where we draw the critical surface which is given
by Eq. \eqref{supcrit}, provided that $\tw(\bar{l})>0$ (see Eq. \eqref{utr}). We also draw the surface
$\tw(\bar{l})=0$, behind which there is the region of the
phase diagram ($\tw(\bar{l})<0$)  that is unaccessible within our RG treatment and where
first-order phase transitions are expected to occur.
The intersection between these two
surfaces gives the TCL at finite temperature $r_{tcl}(T)$, given by Eq. \eqref{TCL}.

In the next section we are going to explore the critical behavior and crossovers around the peculiar second-order phase transition line
$r_{qtc}(T)$
which ends in the QTCP. In this way we will see
how the finite-temperature behavior of observable quantities modifies
due to the influence of the QTCP as a special QCP.

\section{Criticality and crossovers in the influence domain of the quantum tricritical point}
\label{sec3}

According to the conventional RG framework \cite{rudne,ma}, the correlation length $\xi$
and the static susceptibility $\chi\propto \xi^{2-\eta}$
can be expressed as
\bea
\xi=\xi_0 e^{l^*},\qquad \chi=\chi_0 e^{(2-\eta)l^*},
\eea
where $\eta$ is the Fisher exponent and the scale $l^* \gg1$ is determined setting $\tit(l^*)\simeq 1$ \cite{rudne}.
Here, the constants $\xi_0$ and $\chi_0$ are inessential for our purposes, and  will be  included
in the definition of $\xi$ and $\chi$ in the next developments. In our analysis
the Fisher exponent is $\eta=0$.

Here, we  focus on the behavior of susceptibility $\chi$. From  $\tit(l^*)\simeq 1$ and Eq. \eqref{tdielle}
we obtain the self consistent equation for the static susceptibility
\bea
\label{selfcon1}
\chi^{-1}\left[1+K_d \frac{n+8}{4-d}g_{n,d}(T,u) T \chi^{2-\frac{d}{2}}\right]^{\frac{(n+2)}{(n+8)}}\!\!=r -r_c(T),\qquad
\eea
where
\bea
g_{n,d} (T,u) = u-u_{tr}\left(1+\frac{3d-2}{d-2}T^{d-1}\right)\;.
\eea
We consider the peculiar critical line which ends in the QTCP, thus we fix the value of the parameter
$u$ to the coordinate $u_{tr}$ of the QTCP.

The self-consistent equation \eqref{selfcon1} for $\chi$ then becomes
\bea
\label{selfcon2}
\chi^{-1}\left[1+c_{n,d} v T^{d}\chi^{2-\frac{d}{2}}\right]^{\frac{(n+2)}{(n+8)}}=r -r_{qtc}(T),\quad
\eea
where $c_{n,d}=\frac{(n+8)(n+4)(3d-2)K_d^2}{2(d-1)(d-2)(4-d)}$.

If we consider isothermal paths in the ($r,T$)-phase diagram, we obtain two asymptotic regimes for
the static susceptibility, which correspond to mean-field behavior and ``classical Wilsonian" one, namely
\bea
\chi \simeq (r - r_{qtc}(T))^{-\gamma},
\eea
with
\bea
\label{gamma}
\gamma = \left\{ \ba
1\;, & \;{\rm for}\quad r\gg r^{tr}_G(T)\\
1 + \frac{n+2}{2(n+8)}(4-d)\;, & \;{\rm for}\quad r_{qtc}(T) < r\ll r^{tr}_G(T)\;,
\ea \right.
\eea
where
\bea
\label{gi}
r^{tr}_G(T)\simeq r_{qtc}(T)+(c_{n,d} v)^{2/(4-d)} T^{2d/(4-d)}
\eea
is the crossover line which separates these two regimes,
which we call Ginzburg line since it is related to the Ginzburg criterion. To obtain the expected expression \eqref{gamma} for $\gamma$
in the critical or Wilsonian (W)  region, we have used an expansion to first order in $4-d$,
as it is normally done for classical continuous phase transitions.
Thus we correctly recover the critical behavior close to a finite-temperature critical point \cite{noi07,noi10}.
 For conventional QCP, the corresponding Ginzburg line, associated to the generic critical line Eq.\eqref{supcrit}, is given by
(see Eq.\eqref{selfcon1})
\bea
\label{gi2}
r_G(T) &=&
r_{c}(T) + \left\{K_d\frac{n+8}{4-d}
\left[u+\frac{n+4}{2(d-1)}K_d v\Big(1+ \right.\right. \nonumber \\
&+& \left.\left.\left. \frac{3d-2}{d-2}T^{d-1}\right)\right] T \right\}^{2/(4-d)}\;.
\eea

If we approach the critical line fixing the value of the non-thermal parameter $r$ (with $r<r_{tr}$),
decreasing temperature, Eq. \eqref{selfcon2} can be rewritten as 
\bea
&&\chi^{-1}\left[1+c_{n,d} v T^{d}\chi^{2-d/2}\right]^{(n+2)/(n+8)}  \nonumber \\
&&\simeq 2 b_{n,d}\psi T_c(r)^{2\psi-1}(T-T_c(r))\;,
\eea
where we have used, in the right side of Eq.\eqref{selfcon2}, an expansion to first order in $T-T_c(r)$ (i.e. for temperatures close
to the critical one for a given value of $r$).
Then we recover again the classical exponent $\gamma$, which is the same as for isothermal paths (see eqs. \eqref{gamma}) to first order in $4-d$.

The most experimentally interesting situation occurs when we fix both $u$ and $r$ at the QTCP values (Eq. \eqref{QTCP}) and we
study the behavior of susceptibility decreasing temperature. We call this line, with $r=r_{tr}$, ``quantum tricritical
trajectory" in analogy with the usual quantum critical case \cite{noi07}. The self-consistent equation \eqref{selfcon2}
then becomes
\bea
\chi^{-1}\left(1+c_{n,d}v T^{d} \chi^{2-\frac{d}{2}}\right)^{\frac{(n+2)}{(n+8)}}= b_{n,d}v T^{2\psi}.\quad
\eea
At sufficiently low temperature we obtain
\bea
\chi \simeq (b_{n,d} v)^{-1} T^{-2\psi} \sim T^{-2(d-1)}\;,
\eea
which differs from the behavior occurring close to a conventional QCP ($\chi \sim T^{-\psi}$) by the factor 2 in the exponent.
Thus we still have a power-law behavior in temperature when we are in the influence domain of a QTCP, but with a different power
compared to the case of a QCP.

Let us now consider the region of the phase diagram for values $r>r_{tr}$, where the self-consistent equation for $\chi$
is properly written as
\bea
\label{selfcon4}
\chi^{-1}\left[1+c_{n,d} v T^{d}\chi^{2-\frac{d}{2}}\right]^{\frac{(n+2)}{(n+8)}}=r -r_{tr}+b_{n,d}v T^{2\psi}.\qquad
\eea
We  distinguish
the cases in which $(r -r_{tr})$ dominates over the $T$-dependent term on the right side of \eqref{selfcon4}, and viceversa.
So, at sufficiently high temperature, i.e. for $T\gg T^{\dagger \dagger}(r)$, where
\bea
T^{\dagger \dagger}(r) = \left(\frac{r-r_{tr}}{b_{n,d} v}\right)^{1/2\psi},
\eea
the behavior of susceptibility (and hence other observable quantities) is essentially
the same as the one occurring along the quantum tricritical trajectory.

Below this crossover line (i.e. for $T< T^{\dagger \dagger}(r)$) we have to be careful to the presence of the other crossover line \eqref{QR},
 which  separates the classical and quantum regime. Indeed,
 the self-consistent equation obtained for $\chi$ in this section, coming
from the solutions of the RG recursion relations \eqref{rgT} at finite temperature,
is not valid in the quantum regime (i.e. for $T< T^{\dagger}(r)$, which we call Q$_1$ regime).
Hence for $T^\dagger(r) <T<T^{\dagger \dagger}(r)$, we have 
\bea
\chi^{-1} \simeq (r-r_{tr})+b_{n,d}v T^{2\psi}\;,
\eea
where now $r-r_{tr}$ dominates and $b_{n,d}v T^{2\psi}$ represents the leading $T$-dependent
deviation from the ($T=0$) mean-field behavior occurring in the quantum regime. 
Hence, even if we are above $T^\dagger(r)$, we call Q$_2$ this region of temperature .

In the Q$_1$ regime the present formulation misses the exponentially
small correction in temperature (see Ref. \cite{noi07})
to the leading mean field behavior of susceptibility $\chi \sim (r-r_{tr})^{-1}$.

\begin{figure}
  \begin{center}
   \includegraphics[width=8 cm]{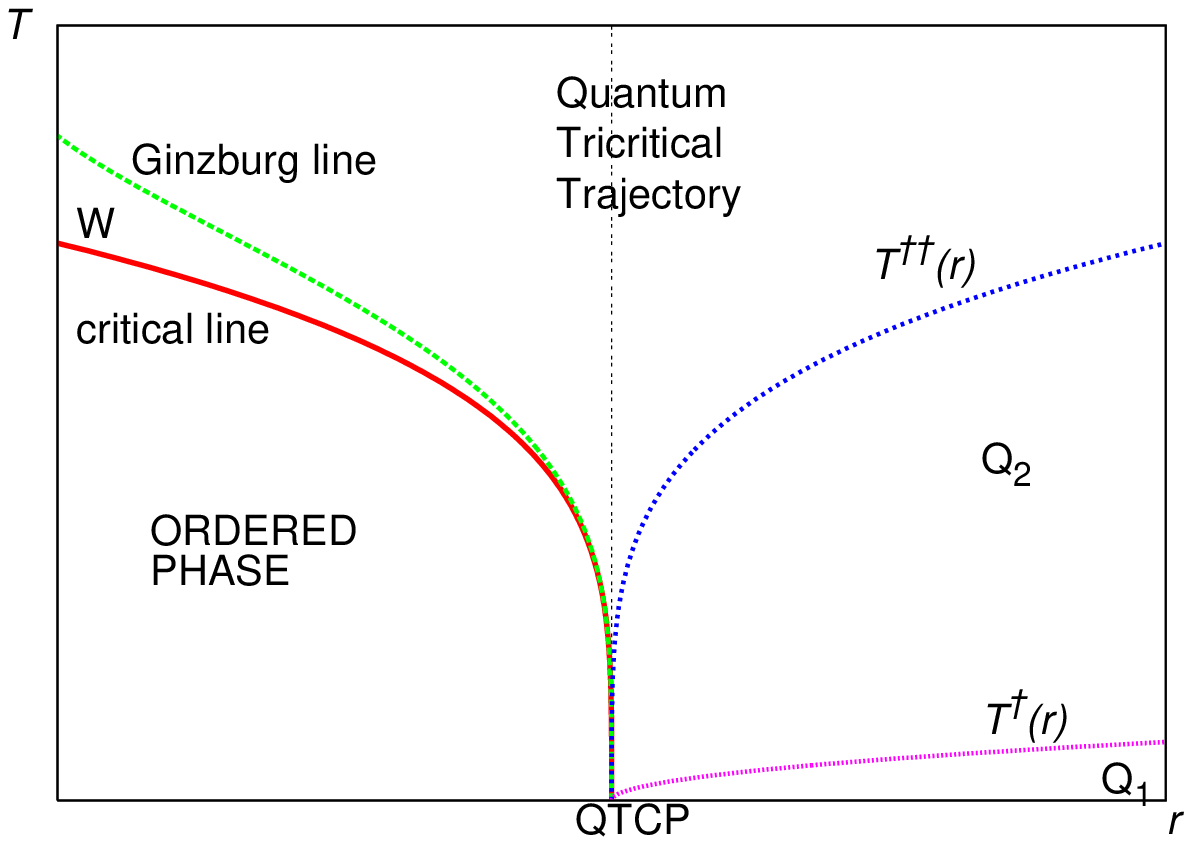}\qquad
   \includegraphics[width=8 cm]{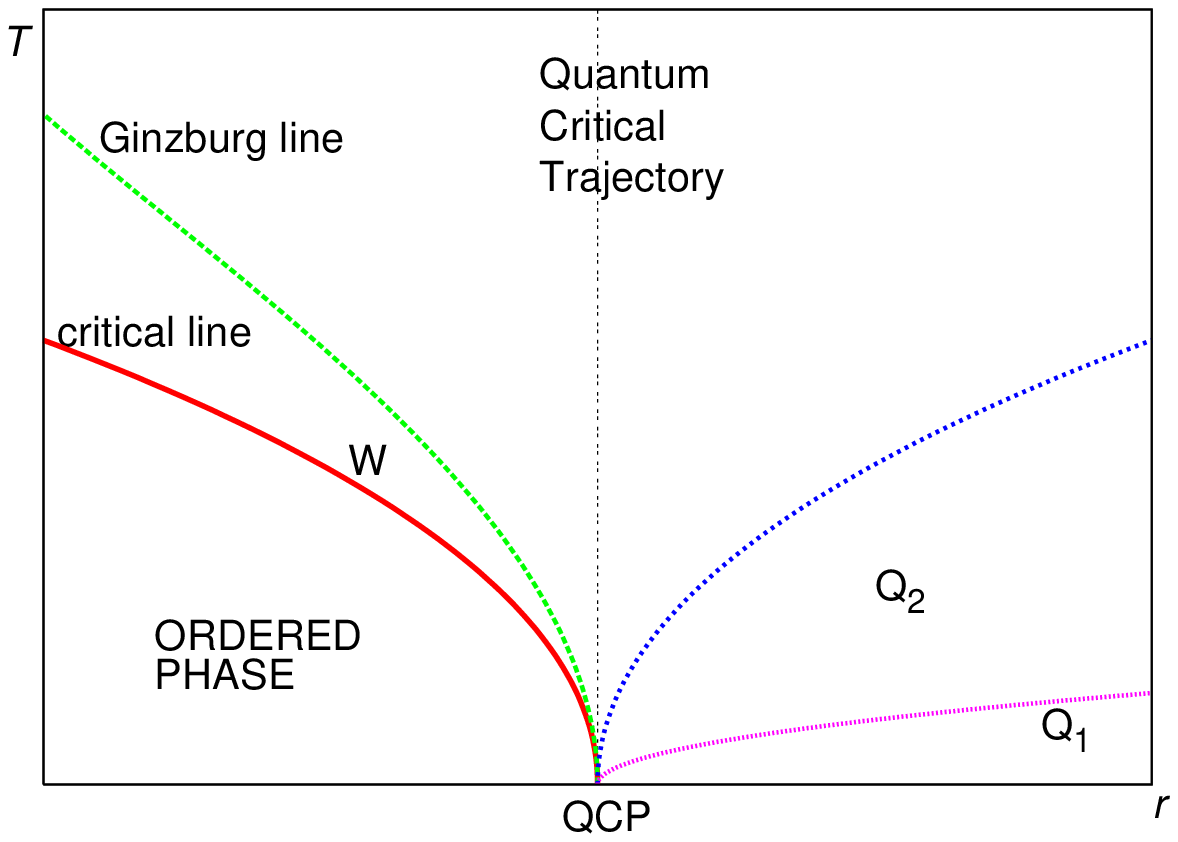}
  \end{center}
  \caption{Low-temperature phase diagram in the ($r,T$)-plane in the presence of a QTCP (top)
  and  when the critical line ends in a QCP (bottom). The nomenclature of the regions is the same
  for both figures:  above the critical line we have the Wilsonian (W) critical region
  where we recover classical critical exponents; on the right side of the phase diagrams we have the ``quantum" regimes Q$_1$ and Q$_2$, where the
  behavior of the susceptibility is essentially mean-field in terms of $(r-r_{tr})$ or $(r-r_c)$ with small corrections in temperature.
  The main differences occur in the fan-shaped region above the QTCP and the QCP, where the static susceptibility behaves as $\chi\sim T^{-2\psi}$
  and $\chi\sim T^{-\psi}$ respectively.   Notice also the different shape
  of the phase boundaries and the size of the Wilsonian regions between the two cases.}
  \label{fig2}
\end{figure}

In Fig. \ref{fig2} we present the low-temperature phase diagram in the ($r,T$)-plane
in the influence domain of a QTCP emerging from our analysis and, for comparison, also
the corresponding phase diagram that we obtain when the critical line ends in an usual QCP.

This figure suggests that the phase diagrams in the $(r,T)$-plane close to a QCP and a QTCP (as a special QCP) appear
very similar and hence hard to distinguish in experiments at very low temperatures. However an observed deviation from the conventional
quantum criticality could be simply explained as a manifestation of the different nature of fluctuations in the influence
domains of QCP and QTCP. Besides, it clearly emerges that, close to the QTCP, the fluctuations are sensibly weaker than those in the influence region
of a conventional QCP.
From Eqs. \eqref{gi} and \eqref{gi2}, we have indeed that the amplitudes of the W-regions are given by
$\Delta_G^{tr}(T)=r_G^{tr}(T)-r_{qtc}(T)\sim T^{2d/(4-d)}$ and $\Delta_G(T)=r_G(T)-r_c(T)\sim T^{2/(4-d)}$ (for $u>0$),
so that one has always $\Delta_G^{tr}< \Delta_G(T)$ for $3\leq d <4$.
This leads to the experimentally interesting conclusion that the quantum tricriticality is well described essentially
in terms of mean field exponents, combined with the shift exponent
$\phi=1/2\psi=1/2(d-1)$, also very close to the full critical line which ends at the QTCP.

\section{Conclusions}
\label{sec4}

In the present paper we have explored the quantum tricriticality,
i.e. the low-temperature properties and crossovers around the
QTCP, of a variety of $d$-dimensional systems with transverse
Ising-like symmetry which could exhibit a QTCP by variation of the
effective coupling parameters in the quantum
Ginzburg-Landau-Wilson action. The study has been performed via a
perturbative RG approach in the Hertz-Millis spirit, which works
very well for dimensionalities $3\leq d<4$, focusing on the shape
of the phase boundaries and on the behavior and crossovers of
susceptibility or correlation length.
It is worth emphasizing that, in this range of dimensionalities,
where tricriticality is governed by a Gaussian fixed point (GFP)
and criticality by the critical FP, the coupling parameter $v$ (or
$\tilde{v}$ in the classical regime), which enters the problem
through the $O(\psi^6)$ term in the quantum action \eqref{S1},
flows towards zero under iteration of the RG transformation and
hence it is an ``irrelevant" variable (in the RG sense).
Nevertheless, it plays a crucial role in determining the correct
physics and it must be necessarily included in the RG analysis
since the initial condition $v>0$ is essential for thermodynamic
stability whenever $u<0$ \cite{aha}. Of course the $O(\psi^6)$
coupling becomes marginal at $d=3$ in the classical regime
implying logarithmic corrections to the Gaussian tricriticality.
We note also that in our RG scenario, as an effect of
fluctuations, the key parameters which govern criticality and
tricriticality (and the related crossovers) are not the original
ones $r$ and $u$ but rather the new scaling fields $t$ and $w$
(here for formal simplicity we refer to $t$ and $w$ both for
classical or quantum regime), with $w>0$ implying second-order
phase transitions and $w<0$ first-order ones. So, tricriticality
is approached when $t$ and $w$ go to zero with $v>0$, by variation
of temperature and the original coupling parameters. In
particular, the tricritical line, ending in the QTCP, is
determined by the equations $t=w=0$ (and not by $r=u=0$ as in the
Landau theory). Remarkably, the inclusion of fluctuations produces
also a shift of the phase boundaries in the global phase diagram
with respect to that resulting  from the Landau theory (see. Fig.
\ref{fig1}).
The emergent quantum tricritical scenario, as compared with that
induced by QCP-fluctuations, provides some peculiar features which
could be very useful in interpreting eventual deviations from the
predictions expected in the influence domain of a conventional
QCP. In particular, we find three key ingredients which may be of
experimental interest. The first one consists in a peculiar phase
diagram where the amplitude of the Wilsonian classical critical
region around the selected critical line ending in the QTCP is
considerably smaller than that occurring near the standard quantum
critical regime. Next, a tricritical shift exponent $\phi^{tr}$ is
derived which is identical to that found in the conventional
quantum criticality. Finally, a peculiar crossover of the critical
shift exponents (which characterize the low-temperature behaviors
of different critical lines) from the conventional value $\phi  =
1/(d-1)$ to the new one  $\phi = 1/2(d-1)$ takes place when the
system under study is tuned between quantum criticality and
quantum tricriticality by variation of the non-thermal control
parameter. It is worth emphasizing again that the QTCP scenario
emerging from our analysis has been obtained for the particular
universality class of $n$-vector quantum systems with TIM-like
dynamics. However, it can be easily extended to include other
quantum materials (as, for instance, the oxides and the
heavy-fermion compounds mentioned in the introduction) where one
expects that a QTCP may be induced by variations of certain
external parameters. For a class of metals, a study of the
crossovers between quantum criticality and quantum tricriticality
has been recently performed \cite{metz} by using a functional RG
approach. However, it invokes the Hertz action and involves
numerous approximations, as the linearization of the flow
equations with the aim to obtain an analytical solution, which are
not easily controllable, especially  in describing crossover
phenomena. This is avoided in our analysis and our predictions
support the use of a direct perturbative RG framework for
obtaining reliable analytical results, at least around realistic
dimensionalities. In this sense, it constitutes again a good
 tool to have a suitable, although qualitative,
scenario concerning the essential physics which occurs around a
conventional QCP and around a QTCP
and the crossovers between these two asymptotic regimes as a
manifestation of the interplay between quantum
critical and tricritical fluctuations.


\end{document}